\documentclass[aps,twocolumn,superscriptaddress]{revtex4-1}
\usepackage[colorlinks=true,bookmarks=false,citecolor=magenta,urlcolor=magenta,linkcolor=magenta,breaklinks]{hyperref} 
\usepackage{breakurl}
\usepackage{sidecap}
\usepackage{amssymb}
\usepackage{hhline}
\sidecaptionvpos{figure}{t}
\usepackage{amsmath}
\usepackage{graphicx}
\usepackage{esint}
\usepackage{epstopdf}
\usepackage{rotating}
\graphicspath{{pict/}{./}}
\usepackage{bm}%
\usepackage{titlesec}

\newcounter{Fig}

\begin{document}

\title{Line Singularities and Hopf Indices of Electromagnetic Multipoles}
\author{Weijin Chen}
\affiliation{School of Optical and Electronic Information, Huazhong University of Science and Technology, Wuhan, Hubei 430074, P. R. China}
\author{Yuntian Chen}
\email{yuntian@hust.edu.cn}
\affiliation{School of Optical and Electronic Information, Huazhong University of Science and Technology, Wuhan, Hubei 430074, P. R. China}
\author{Wei Liu}
\email{wei.liu.pku@gmail.com}
\affiliation{College for Advanced Interdisciplinary Studies, National University of Defense
Technology, Changsha, Hunan 410073, P. R. China}
\begin{abstract}
Electromagnetic multipoles can be continuously mapped to tangent vectors on the momentum sphere, the topology of which guarantees the existence of isolated singularities. For pure (real or imaginary) vectors, those singularities correspond to zeros of tangent fields, which can be classified by integer Poincar\'{e} indices. Nevertheless, electromagnetic fields are generally complex vectors, a comprehensive characterization of which requires the introduction of line fields and line singularities categorized by half-integer Hopf indices. Here we explore complex vectorial electromagnetic multipoles from the perspective of line singularities, focusing on the special case of polarization line field. Similar to the case of pure vectors, the Poincar\'{e}-Hopf theorem  forces the index sum of all line singularities to be $\mathbf{2}$, irrespective of the specific multipolar compositions. With this multipolar insight, we further unveil the underlying structures of radiative circularly-polarized Bloch modes of photonic crystal slabs, revealing their topological origins with line singularities of Hopf indices. Our work has established subtle connections between three seemingly unrelated but sweeping physical entities (line singularities of Hopf indices, electromagnetic multipoles, and Bloch modes), which can nourish new frames of visions and applications fertilizing many related fields.
\end{abstract}
\maketitle

Presaged by early works of Dirac, Van Hove, Aharanov and Bohm, concepts of geometry and topology have currently become essential in almost every branch of physics and many other disciplines~\cite{NING_2013__Selected,YANG_2018__Topology,DAVID_1998__Topological,BERRY_2017__HalfCentury,ELSDALE_1976_WilhelmRouxsArch.Dev.Biol._Fibroblast,PENROSE_1979_Ann.Hum.Genet._topology}. Spurred especially by explosively expanding fields of topological insulators and semimetals~\cite{BERNEVIG_2013__Topological,ARMITAGE_2018_Rev.Mod.Phys._Weyl}, topological concepts have now rapidly pervaded the classical fields of photonics, mechanics and acoustics, bring unprecedented new vistas for manipulations of photons, phonons, and particle-waves of other forms~\cite{Lu2014_topological,OZAWA_2018_ArXiv180204173,YANG_Phys.Rev.Lett._topological_2015-1,HUBER_2016_Nat.Phys._Topological}. Recently a topological perspective has been incorporated into another sweeping concept of electromagnetic multipoles~\cite{jackson1962classical,Bohren1983_book}, treating the isolated dark directions of radiations as vectorial singularities and assigning each of them a Poincar\'{e} index~\cite{CHEN_2019__Singularities}. This Poincar\'{e}'s vision for Mie's multipoles has constructed a new bridge to another seemingly unrelated field of bound states in the continuum (BICs)~\cite{HSU_Nat.Rev.Mater._bound_2016}, establishing naturally a lucid connection between singularity indices and topological charges of those BICs~\cite{HSU_Nature_observation_2013-1,ZHEN_2014_Phys.Rev.Lett._Topological}.

The Poincar\'{e} indices and  original Poincar\'{e}-Hopf theorem can only be directly applied to pure (real or imaginary) tangent vectors on manifolds~\cite{MILNOR_1997__Topology}. Nevertheless, electromagnetic waves generally are complex vectors, which alternatively can be viewed as a combination of two sets of pure vectors (real and imaginary parts of the fields). For each set, the vectorial version of Poincar\'{e}-Hopf theorem with integer Poincar\'{e} indices can be applied, but this approach cannot fully grasp electromagnetic fields by its complex vectorial nature. A full characterization lies in the derivative line field (or equivalently a tensor field~\cite{PENROSE_1979_Ann.Hum.Genet._topology,DELMARCELLE_1994__visualization,BISHOP_2012__Tensor}), which dates back to the original seminal  work of Hopf~\cite{HOPF_2003__Differential}). For line fields an analogues Poincar\'{e}-Hopf theorem can be applied to its line singularities of half-integer Hopf indices~\cite{HOPF_2003__Differential,DELMARCELLE_1994__visualization}. The construction of line fields from complex vectorial fields is not unique, whereas their Hopf index sum is decided only by the topology of the manifold it is mapped to, regardless of the specific form of the line field introduced~\cite{BOSCAIN_2016_DifferentialGeometryanditsApplications_Generic,CROWLEY_2017_JournalofGeometryandPhysics_Poincare}. The most widely adopted special case of line fields for electromagnetic waves is the polarization field, of which the line (polarization) singularities are either \textbf{V}-points (where the  field intensity is zero) or \textbf{C}-points (where the field is circularly-polarized)~\cite{BERRY_2017__HalfCentury,NYE_natural_1999,DENNIS_2009_ProgressinOptics_Chapter,GBUR_2016__Singular,FREUND_2002_Opt.Commun._Polarization}.

Here we reexamine electromagnetic multipoles of general complex vectorial forms, focusing on their topological properties of polarization singularity distributions and their indices. Both individual multipoles and their combinations are studied, confirming the Poincar\'{e}-Hopf theorem for line fields, which requires that the index sum across the momentum sphere is always $\mathbf{2}$. We have achieved not only the lowest-order singularities (single-twist M\"{o}bius strips) of index $\pm 1/2$, but also higher-order counterparts (multi-twist M\"{o}bius strips) with larger half-integer indices. This Hopf's vision of multipoles enables us to trace the origin of radiative circularly-polarized Bloch modes to the overlapping of polarization singularities with open radiation channels of photonic crystal slabs. Equivalence between singularity Hopf indices and topological charges of those Bloch modes are also revealed. Our approach of connecting both local and global topological properties of electromagnetic multipoles can be simply extended to propagating vectorial vortex beams of various spatial dependencies and polarizations, which can significantly expand the horizons of the broad field of singular photonics~\cite{GBUR_2016__Singular,Rosales_Guzm_n_2018}.

\begin{figure}[tp]
\centerline{\includegraphics[width=8.5cm]{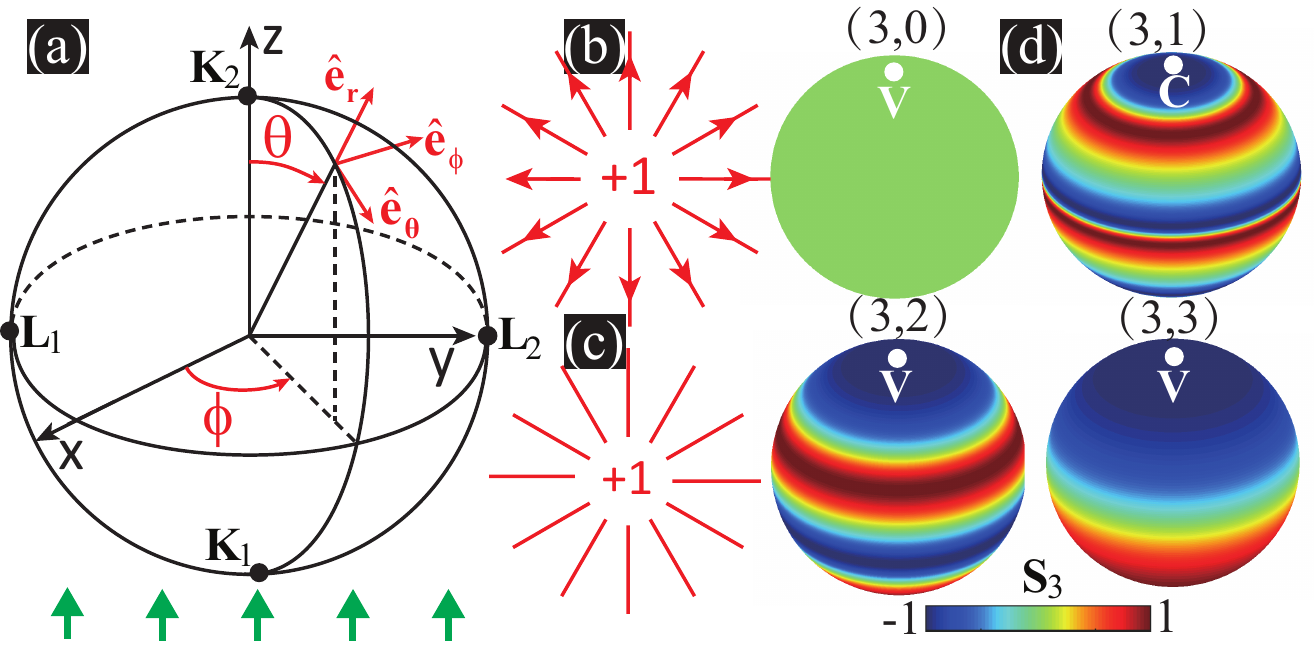}} \caption{\small (a) Cartesian and polar coordinate systems  with the associated orthonormal basis vectors $\mathbf{\hat{e}}_\theta$, $\mathbf{\hat{e}}_\phi$ and $\mathbf{\hat{e}}_r$ indicated. An plane wave (green-arrow array) is propagating along $\mathbf{z}$ direction. (b) and (c): vector and line patterns around the orientable singularity of index $+1$. (d) Polarization distributions (in terms of $S_3$) for individual multipoles of order $(l,m)$, with singularities locating at both poles of index $+1$ (only those on the north poles are indicated).}
\label{fig1}
\end{figure}

To illustrate the necessity to introduce line fields for the characterization of complex vectorial fields, in Fig.~\ref{fig1}(a) we show the most elementary case of mapping a plane wave (propagating along $\mathbf{z}$ direction) to \textit{tangent vectors} on the unit sphere. When the plane wave is linearly-polarized, the tangent vectors are pure with two singularities [zeros of tangent fields; each with a Poincar\'{e} index $\mathbf{Ind}_p=1$, with the field patterns around shown in Fig.~\ref{fig1}(b)] located on the equator: \textit{e.g.} for $\mathbf{y}$-polarized waves, points $\textbf{L}_{1,2}$ indicated are tangent-field singularities. For pure vectorial field, a line field can be directly obtained with the arrows of vectors removed~\cite{BOSCAIN_2016_DifferentialGeometryanditsApplications_Generic,CROWLEY_2017_JournalofGeometryandPhysics_Poincare}. For this natural definition, the line singularities coincide with the vector singularities, which are also termed as orientable singularities with the Hopf indices $\mathbf{Ind}_h=\mathbf{Ind}_p$~\cite{BOSCAIN_2016_DifferentialGeometryanditsApplications_Generic,CROWLEY_2017_JournalofGeometryandPhysics_Poincare}. The situation would become contrastingly different when the plane wave is generally elliptically-polarized, with no \textbf{V}-points of tangent fields on the sphere. Of course we can always decompose the wave into two sets of linearly-polarized ones that correspond to pure vectors, but this is not sufficient for a full characterization of the complex vectors.  Alternatively, we can  introduce a tangent polarization line field~\cite{BERRY_2017__HalfCentury,NYE_natural_1999,DENNIS_2009_ProgressinOptics_Chapter,GBUR_2016__Singular,FREUND_2002_Opt.Commun._Polarization} and then locate the line singularities and identify their Hopf indices: \textit{e.g.} for circularly-polarized plane waves, both north and south poles (indicated by $\textbf{K}_{1,2}$) are singularities [\textbf{C}-points of Hopf index $\mathbf{Ind}_h=+1$, with the line patterns around shown in Fig.~\ref{fig1}(c)]. The line patterns shown correspond to polarization lines in terms of semi-major axis of the polarization ellipses, as is the case throughout this work.

Now we turn to complex electromagnetic multipoles, the fundamental significance of which lies in the fact that they constitute a complete orthogonal basis for expansions of arbitrary electromagnetic radiations~\cite{jackson1962classical,Bohren1983_book,GRAHN_NewJ.Phys._electromagnetic_2012,Supplemental_Material}: ${{\bf{E}}_{\rm{rad}}}\left( {r,\theta ,\phi } \right) = \sum\nolimits_{l = 1}^\infty {\sum\nolimits_{m =  - l}^l {{E_{lm}}\left[ {{a_{lm}}{{\bf{{N}}}_{lm}} + {b_{lm}}{{\bf{{M}}}_{lm}}} \right]} }$. Here ${E_{lm}} = {i^{l + 1}}\left( {2l + 1} \right)\sqrt {{{\left( {l - m} \right)!} \mathord{\left/
 {\vphantom {{\left( {l - m} \right)!} {4l\left( {l + 1} \right)\left( {l + m} \right)!}}} \right.
 \kern-\nulldelimiterspace} {4l\left( {l + 1} \right)\left( {l + m} \right)!}}} $; and the complex spherical harmonics $\bf{{N}}$$_{lm}$ and $\bf{{M}}$$_{lm}$ that correspond respectively to electric and magnetic multipoles are~\cite{jackson1962classical,Bohren1983_book,GRAHN_NewJ.Phys._electromagnetic_2012,Supplemental_Material}:
\begin{equation}
\label{spherical_harmonics}
\begin{aligned}
{{{\bf{{N}}}}_{lm}}{\rm{ = }}\left[ {{\tau _{lm}}\left( {\theta } \right){{{\bf{\hat e}}}_\theta } + i{\pi _{lm}}\left( {\theta } \right){{{\bf{\hat e}}}_\phi }} \right]{{{{\left[ {kr{z_l}} \right]}^\prime }} \over {kr}}\exp \left( {im\phi } \right);\\
{\mathbf{{M}}_{lm}} = \left[ {i{\pi _{lm}}\left( {\theta } \right){{{\bf{\hat e}}}_\theta } - {\tau _{lm}}\left( {\theta } \right){{{\bf{\hat e}}}_\phi }} \right]{z_l}\exp \left( {im\phi } \right),
\end{aligned}
\end{equation}
 where we have dropped the terms along ${{{\bf{\hat e}}}_r}$ for $\bf{{N}}$$_{lm}$ as we are only interested in tangent fields; ${\tau _{lm}}\left( \theta  \right) = {d \over {d\theta }}P_l^m\left( {\cos \theta } \right)$; ${\pi _{lm}}\left( \theta  \right) = {m \over {\sin \theta }}P_l^m\left( {\cos \theta } \right)$; $P_l^m\left( {\cos \theta } \right)$  denotes associated Legendre polynomials; and ${{z_l}\left( {kr} \right)}$ is spherical Bessel or Hankel function depending on the radiation type (\textit{e.g.} ${{z_l}}$ would correspond to the spherical Hankel function of the first and second kinds for outgoing and incoming radiations, respectively)~\cite{BRONSHTEIN_2007__Handbook}. Here in Eq.~(\ref{spherical_harmonics}) we have adopted complex notations, rather than real ones employed in Ref.~\cite{CHEN_2019__Singularities}, where all singularities of an individual multipole are orientable, with equal Poincar\'{e} and Hopf indices.

 As a first step, we will locate the singularities of an individual multipole of order $(l,m)$. The polarization of each point on the momentum sphere is decided only by the quotient $Q_{lm}(\theta)={\pi _{lm}}(\theta)/{\tau _{lm}}(\theta)$, which is $\phi$-independent. As a result, the \textit{isolated singularities} can only located on the poles where $\phi$ is not defined. At the same time, the Poincar\'{e}-Hopf theorem guarantees the existence of isolated singularities with index sum of $\mathbf{2}$ and moreover the two poles are symmetric. This enables us to draw the conclusion, that for an individual multipole of arbitrary order, there are only two isolated polarization singularities with indices:
\begin{equation}
\label{index_ns}
{\mathbf{Ind}_h}=+1,~~{|\cos \theta|=1}.
\end{equation}
Considering that the singularities are non-vanishing only when $m=\pm 1$~\cite{jackson1962classical,Bohren1983_book,CHEN_2019__Singularities}, the poles are \textbf{C}-points for  $m=\pm 1$ and \textbf{V}-points for $m\neq\pm 1$.

To further describe the detailed local polarization distributions,  we can employ the widely adopted normalized Stokes parameters~\cite{Yariv2006_book_photonics,Supplemental_Material}. Since here we focus on singularities, we employ $S_3$: $S_3=\pm 1$ corresponds respectively to left-handed and right-handed circularly-polarized light, which is the necessary condition for an isolated \textbf{C}-point; $S_3=0$ corresponds to either linear polarization or a vanishing radiation, the latter of which is necessary  for an isolated \textbf{V}-point~\cite{Yariv2006_book_photonics,Supplemental_Material}. For electric and magnetic multipoles of the same order, since $\textbf{N}_{lm}\cdot\textbf{M}_{lm}=0$, they can transform to each other by a $90^\circ$ rotation around ${{{\bf{\hat e}}}_r }$~\cite{jackson1962classical,Bohren1983_book,CHEN_2019__Singularities}, and thus the polarization ellipses can be converted to each other by an interchange of the semi-major and semi-minor axes, without flipping the handedness: $S_3(\textbf{N}_{lm})=S_3(\textbf{M}_{lm})$. At the same time, since $P_{\ell}^{-m}(\theta)=(-1)^{m} \frac{(\ell-m) !}{(\ell+m) !} P_{\ell}^{m}(\theta)$ and thus $Q_{lm}=-Q_{l,-m}$, we obtain $S_3(\textbf{M}_{lm}; \textbf{N}_{lm})=-S_3(\textbf{M}_{l,-m}; \textbf{N}_{l,-m})$, indicating the same polarization distribution except for the opposite handedness~\cite{Supplemental_Material}. In Fig.~\ref{fig1}(d) we show the polarization distributions (across the momentum sphere in terms of $S_3$) with the singularities indicated on the momentum sphere for multipoles of order $(l=3,~m=0\rightarrow3)$, and more scenarios are shown in Ref.~\cite{Supplemental_Material}.  As is shown, when $m=0$, it is linearly-polarized throughout the momentum sphere ($S_3=0$). This is due to that ${\pi _{l,m=0}}(\theta)=0$, which has left only one non-vanishing field component [see Eq.~(\ref{spherical_harmonics})] of linear polarization always.
\begin{figure}[tp]
\centerline{\includegraphics[width=8.9cm]{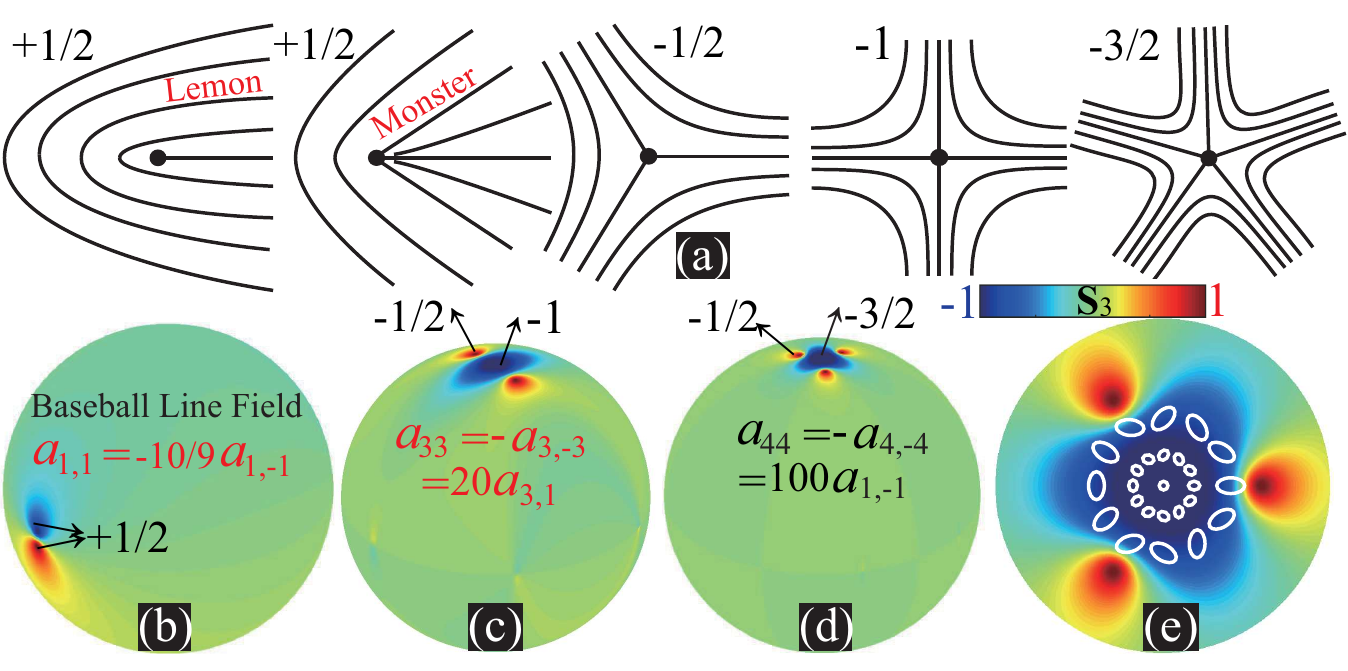}} \caption{\small (a) Elementary line patterns around singularities of different indices, except for the case of $+1/2$ for which a non-elementary pattern of monster is also shown. (b)-(d) Polarization distributions for different combinations of multipoles, where \textbf{C}-points of indices of $\pm 1/2$, $-1$ and $-3/2$ emerge. Detailed polarization ellipses around the singularity of index $-3/2$ are shown in (e).}
\label{fig2}
\end{figure}

For line fields the \textit{fundamental} and stable singularities are the lowest-order ones with ${\mathbf{Ind}_h}=\pm1/2$ (single-twist M\"{o}bius strips~\cite{FREUND_2011_OpticsCommunications_Mobius,BAUER_Science_observation_2015,BAUER_2016_Phys.Rev.Lett._Optical}). Higher-order singularities with larger indices (multi-twist M\"{o}bius strips~\cite{FREUND_2011_OpticsCommunications_Mobius}) are not stable and can be easily broken into to a series of lower-order and/or fundamental ones~\cite{HOPF_2003__Differential,DELMARCELLE_1994__visualization}. To obtain singularities of ${\mathbf{Ind}_h}\neq+1$ (different from those in Fig.~\ref{fig1} of an individual multipole), we can combine multipoles of different orders and natures. In principle, an arbitrary half-integer index can be achieved through the multipolar combinations, as $\textbf{N}_{lm}$ and $\textbf{M}_{lm}$ constitute a complete bases for field expansions.  Figure~\ref{fig2}(a) shows the line patters close to singularities of different indices, where the singularity of index $-1$ is orientable, with line patterns being the same as those of pure vectors with the arrows removed~\cite{CHEN_2019__Singularities,HOPF_2003__Differential,DELMARCELLE_1994__visualization}. It is worth emphasizing that even for singularities of a fixed index, there are infinitely many different sets of associated line patterns, though they are topologically equivalent and can be continuously mapped to one another~\cite{HOPF_2003__Differential,DELMARCELLE_1994__visualization,GALVEZ_2014_Phys.Rev.A_Generation,KHAJAVI_2016_J.Opt._Highorder}. For each index there is the simplest \textit{elementary} pattern: the orientations of the line fields rotate linearly with the traversing angle along a circuit around the singularity. For example, in Fig.~\ref{fig2}(a) for $\mathbf{Ind}_h=+1/2$ we show two line patterns, a lemon and a monster, where the former is elementary while the latter is not~\cite{BERRY_2017__HalfCentury,NYE_natural_1999,DENNIS_2009_ProgressinOptics_Chapter,GBUR_2016__Singular,FREUND_2002_Opt.Commun._Polarization}. The lemon and monster are topologically the same and smoothly inter-convertible~\cite{KUMAR_2015__Monstar} (as is also the case for their higher-order counterparts~\cite{KHAJAVI_2016_J.Opt._Highorder}). The other three line patterns shown in Fig.~\ref{fig2}(a) are all elementary.

For an arbitrary combinations of multipoles, the positions of polarization singularities can be directly mapped from the $S_3$ distributions~\cite{Supplemental_Material}. To obtain the index of each singularity, the complex vectorial field  can be converted to a complex scalar field, of which the dislocation strength is twice the Hopf index~\cite{BERRY_2004_J.Opt.PureAppl.Opt._Index}.  Alternatively and equivalently, on the tangent plane close to the isolated singularity, a local polar coordinate $(r_s, \phi_s)$ (singularity at $r_s=0$) can be introduced.  In this basis, the neighbouring complex vectors can be expressed as~\cite{FREUND_2011_OpticsCommunications_Mobius,GALVEZ_2014_Phys.Rev.A_Generation,KHAJAVI_2016_J.Opt._Highorder}:
\begin{equation}
\label{complex_vector}
\mathbf{E}(r_s, \phi_s)=c_R\mathbf{\hat{e}}_{{R}}+c_{L}\mathbf{\hat{e}}_{{L}},
\end{equation}
where $c_R=\mathcal{R}(r_s)\left(\cos \alpha \mathrm{e}^{\mathrm{i} \ell_{1} \phi_s}+\sin \alpha \mathrm{e}^{-\mathrm{i} \ell_{1} \phi_s} \mathrm{e}^{\mathrm{i} \beta_1}\right)$; $c_L=\mathcal{L}(r_s)\mathrm{e}^{\mathrm{i} \ell_{2} \phi_s} \mathrm{e}^{\mathrm{i} \beta_2}$; $\alpha$ and $\beta_{1,2}$ are phases; $\ell_{1,2}$ are integers; and $\mathbf{\hat{e}}_{{L,R}}$ are left-handed and right-handed circularly-polarized local basis vectors. The singularity at the origin is either a \textbf{V}-point [$\mathcal{R}(0)=\mathcal{L}(0)=0$] or \textbf{C}-point [$\mathcal{R}(0)\cdot \mathcal{L}(0)=0$ and $\mathcal{R}(0)\neq \mathcal{L}(0)$]. Then we can introduce a complex quantity $w=c_R/c_L$ that decides the Hopf index and the neighbouring polarization distributions, since it is directly related to the Stokes vector \textbf{S} on the Poincar\'{e} sphere through the stereographic projection~\cite{PENROSE_2004__Road,BERRYM.V._2003_Proceeding_optical} (refer to Ref.~\cite{Supplemental_Material} for more details). Equivalently, $w$ can be viewed as a $1$-spinor, which is a square root of the associated Stokes vector~\cite{HLADIK_1999__Spinors,FARMELO_2011__strangest}. Equation~(\ref{complex_vector}) encompasses both elementary and non-elementary polarization line patterns, and the Hopf index is dependent on all parameters listed above, and thus can be obtained only case by case through the complex scalar field approach~\cite{BERRY_2004_J.Opt.PureAppl.Opt._Index}. Nevertheless, for singularities with elementary line patterns, or sufficiently close to the singularity,  $w$ can be simplified as \cite{BERRYM.V._2003_Proceeding_optical,Supplemental_Material}:
\begin{equation}
\label{spinor}
w=\rho(r_s)\exp(i\ell\phi_s),
\end{equation}
where $\rho(r_s=0)=0$ or $\infty$ and $\ell$ is an integer. As a circuit is traversed around the singularity (with $\phi_s=0 \rightarrow 2\pi$), the corresponding Stokes vector \textbf{S} would trance out a complete latitude circle $|\ell|$ times, indicating $|\ell|/2$ rounds of $2\pi$ full rotations of the polarization lines along the circuit (refer to Ref.~\cite{Supplemental_Material} for more details). The positive (negative) sign of $\ell$ means that rotation of the line is in the same (opposite) sense as that the contour is traversed. As a result, the index of the singularity at the origin is  ${\mathbf{Ind}_h}(r_s=0)=\ell/2$.

Knowing how to locate the singularities and then calculate their indices, now we can combine different multipoles to obtain singularities required. The simplest scenario for the lowest-order line singularity distributions  would be four singularities across the momentum sphere with equal index of $+1/2$ (also termed as the baseball line field~\cite{BOSCAIN_2016_DifferentialGeometryanditsApplications_Generic}), which makes an index sum of $\mathbf{2}$.  This fundamental type of line field can be realized by breaking two \textbf{V}-points of index $+1$ of the pure vectorial dipole into four \textbf{C}-points, as is shown in Fig.~\ref{fig2}(b). This line field can be directly achieved with scattering particles~\cite{GARCIA-ETXARRI_2017_ACSPhotonics_Opticala}, while unfortunately in previous studies it had not been placed in the broad background of line singularities and Hopf indexes as we have done here, and thus its global properties in terms of index sum had not been revealed. Two extra scenarios are also shown (refer to Ref.~\cite{Supplemental_Material} for more details): in Fig.~\ref{fig2}(c) \textbf{V}-points ($\mathbf{Ind}_h=-2$) of the pure vectorial octupole are broken into three \textbf{C}-points (one of index $-1$ and the other two of index $-1/2$); and  in Fig.~\ref{fig2}(d) \textbf{V}-points ($\mathbf{Ind}_h=-3$) of the pure vectorial hexadecapole ($l=4$) are broken into four \textbf{C}-points  (one of index $-3/2$ and the other three of index $-1/2$). The introduction of the $m=1$ term guarantees that the singularities at the poles are \textbf{C}-points. The polarization ellipses around to the singularity of index $-3/2$ are shown in Fig.~\ref{fig2}(e). In Figs.~\ref{fig2}(b)-(d) not all singularities are shown, but it is easy to confirm that the topological charge is locally conserved during singularity breaking~\cite{Supplemental_Material} and thus global index sum of all singularities is always $\mathbf{2}$, consistent with the Poincar\'{e}-Hopf theorem.

After having thoroughly surveyed the hidden landscapes of electromagnetic multipoles of complex vectorial natures, now we aim to show how their topological properties could be wielded to bring new frames of visions. It is previously shown that periodic photonic structures could be treated in a reductionist manner: the far-field properties of the radiative Bloch modes are decided only by the multipolar radiations of unit-cells along the corresponding directions of open out-coupling channels~\cite{CHEN_2019__Singularities,LIU_2017_ACSPhotonics_Beam,LIU_2018_Opt.Express_Generalized}.
For example, the coincidence of \textbf{V}-points of unit-cell radiations with the open channels would immediately produce BICs, as there is no effective energy leakage~\cite{CHEN_2019__Singularities,HSU_Nat.Rev.Mater._bound_2016}. From this perspective, it is natural to expect that when the multipolar \textbf{C}-points overlap with the open channels, the corresponding Bloch modes would be circularly-polarized in the far-field, and the \textbf{C}-point Hopf indices would decide the corresponding topological charges of those Bloch modes.

\begin{figure}[tp]
\centerline{\includegraphics[width=8.9cm]{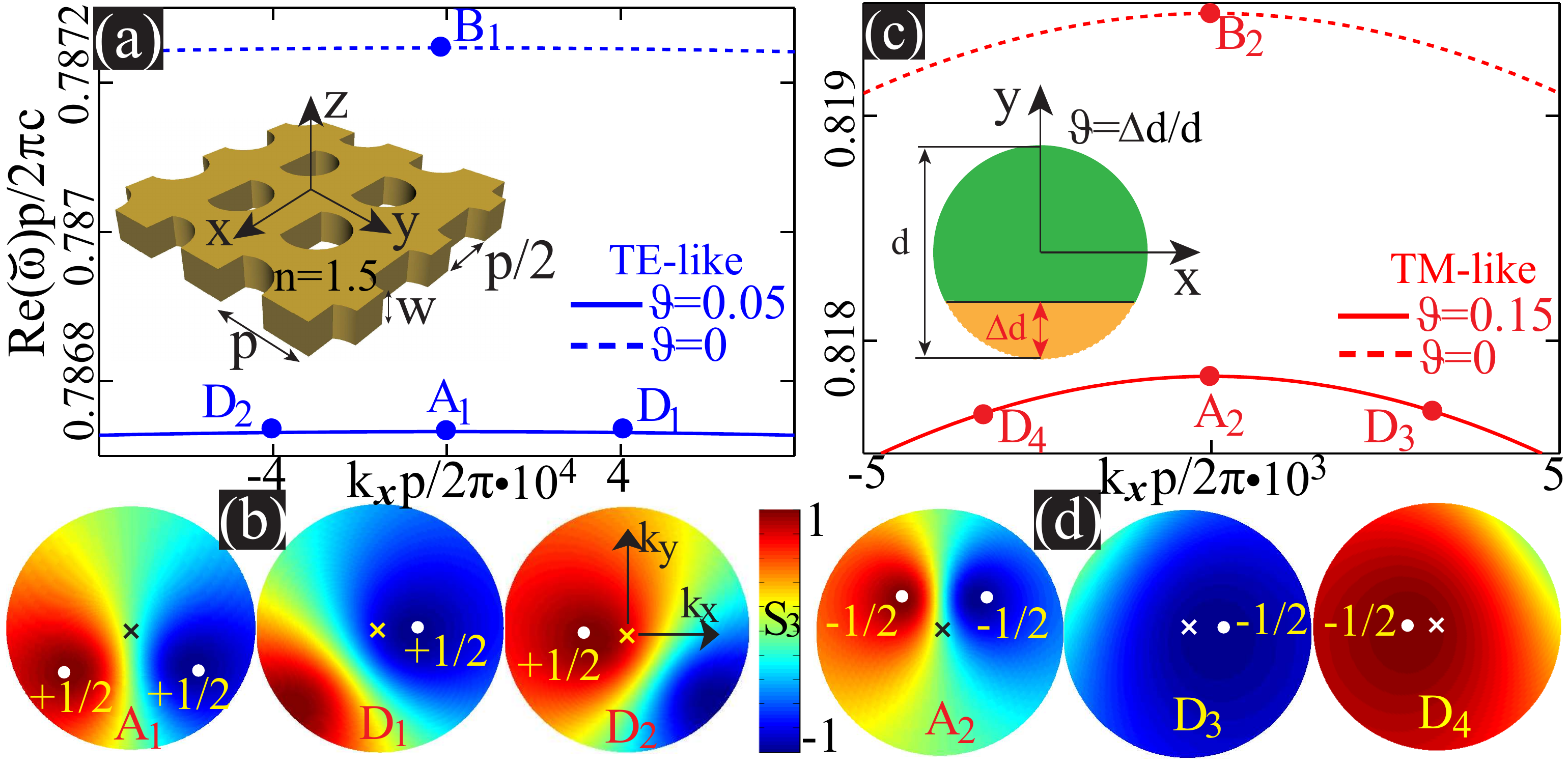}} \caption{\small \small (a) and (c): Dispersion curves of two bands for square photonic crystal slabs (schematically shown as insets) and eight Bloch modes are indicated: $\textbf{A}_{1,2}$ (linearly-polarized); $\textbf{B}_{1,2}$ (BICs);  $\textbf{D}_{1\rightarrow4}$ (circularly-polarized). (b) and (d): corresponding 2D polarization distributions (around the pole of $k_x=k_y=0$ denoted by crosses) for points indicated in (a) and (c), with \textbf{C}-points (denoted by crosses) and their indices specified.}
\label{fig3}
\end{figure}

To confirm what is mentioned above, we turn to photonic crystal slabs with broken symmetries~\cite{KOSHELEV_2018_Phys.Rev.Lett._Asymmetrica} that support such circularly-polarized Bloch modes. We will adopt the technical approach implemented in Ref.~\cite{LIU_2019_ArXiv190401733Phys._Circularly} of breaking BICs into Bloch modes of circular polarizations, but emphasize that our multipolar interpretations and the intuitive physical pictures obtained would be fully exclusive.  Firstly we break the symmetry of a photonic crystal slab with square lattices of circular air holes (slab refractive index $n=1.5$; height $h=p/2$; lattice constant $p$; air hole diameter $d=p/2$) by filling in part of the holes, as are shown in the insets of Figs.~\ref{fig3}(a) and (c), with a filling parameter $\vartheta$ defined. The dispersion curves (in terms of complex eigenfrequencies $\breve{\omega}$~\cite{HSU_Nature_observation_2013-1,ZHEN_2014_Phys.Rev.Lett._Topological,CHEN_2019__Singularities} calculated through COMSOL Multiphysics; $c$ is the speed of light) of two bands (TE-like and TM-like) are shown in Fig.~\ref{fig3}(a) and (c), where for better comparison the dispersion curves of the symmetric case ($\vartheta=0$) are also shown. As has been revealed in~\cite{CHEN_2019__Singularities}, the indicated points $\textbf{B}_{1,2}$ denote $\Gamma$-point BICs with topological charges of $\pm 1$, which correspond to \textbf{V}-points of multipolar radiations with Hopf indices of $\pm 1$, respectively.

\begin{figure}[tp]
\centerline{\includegraphics[width=8.5cm]{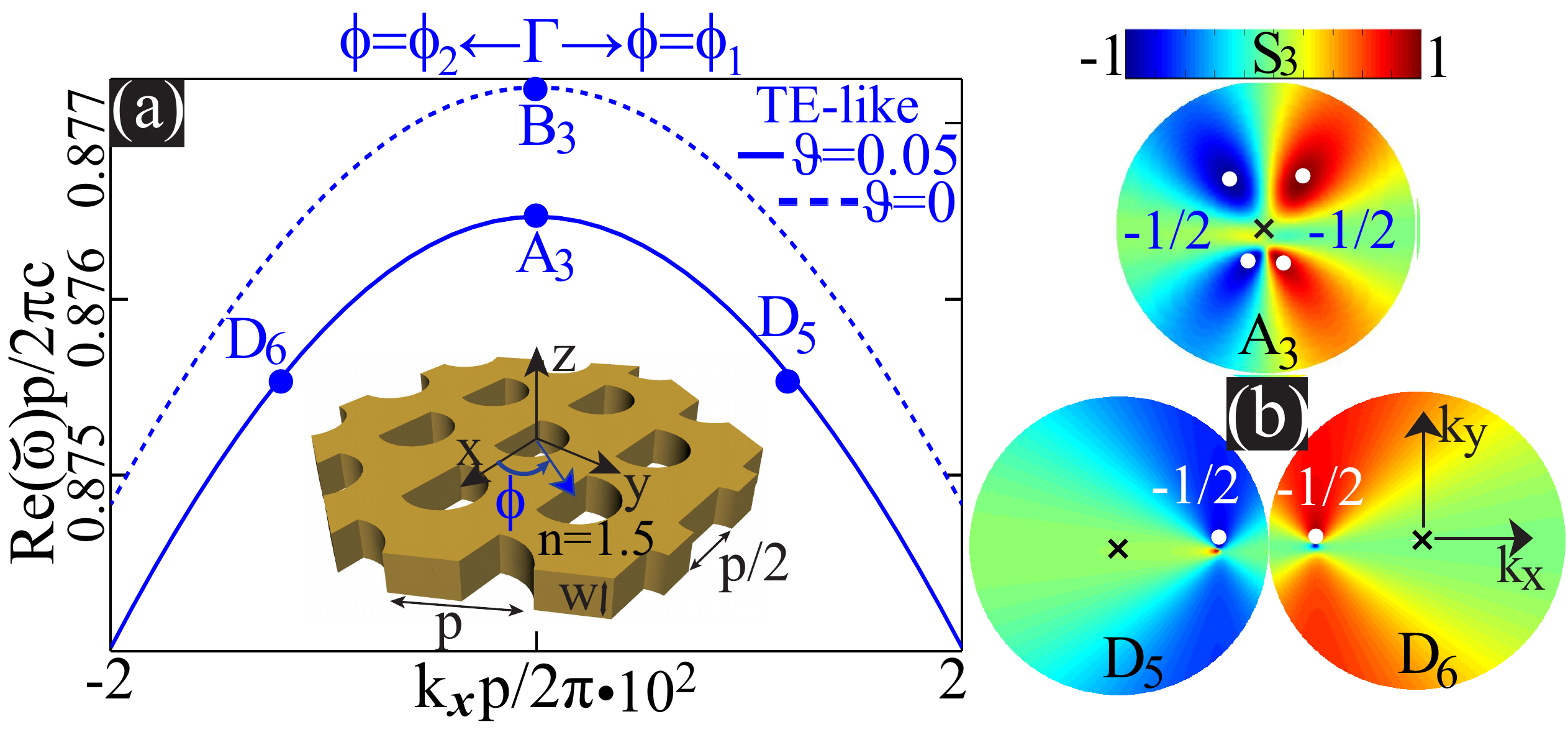}} \caption{\small (a): Dispersion curves for hexagonal slabs (see the inset) and four Bloch modes are indicated: $\textbf{A}_{3}$ (linearly-polarized); $\textbf{B}_{3}$ (BIC);  $\textbf{D}_{5,6}$ (circularly-polarized). The corresponding 2D polarization distributions are shown in (b), with \textbf{C}-points and their indices specified.}
\label{fig4}
\end{figure}

When the symmetry is broken ($\vartheta\neq0$),  similar to the scenarios discussed in Fig.~\ref{fig2}, at $\Gamma$-points ($\textbf{A}_{1,2}$) the \textbf{V}-point of the multipolar radiation (refer to Ref.~\cite{Supplemental_Material} for more details of multipolar expansions and specific multipolar compositions at those points indicated in Figs.~\ref{fig3} and \ref{fig4}) will break into a pair of singularities of indices $+1/2$ [$\textbf{A}_{1}$ in Fig.~\ref{fig3}(b) with $\vartheta=0.05$] or $-1/2$ [$\textbf{A}_{2}$ in Fig.~\ref{fig3}(d) with $\vartheta=0.15$], both with opposite handedness.  At $\textbf{A}_{1,2}$, along the open channel ($k_x=k_y=0$) the multipolar radiation is linearly-polarized ($S_3$=0) [Figs.~\ref{fig3}(b) and (d)], indicating that the far-field radiations of the Bloch modes at $\Gamma$-point are also linearly-polarized (not BIC anymore). To obtain Bloch modes of circular polarizations, overlapping of the C-points with the open channels are required, which is satisfied at the points indicated in Fig.~\ref{fig3}(a) ($\textbf{D}_{1,2}$: $k_xp/2\pi=\pm 0.0004$ and $k_y=0$) and Fig.~\ref{fig3}(b) ($\textbf{D}_{3,4}$: $k_xp/2\pi=\pm 0.0033$ and $k_y=0$).  We note here that such circularly-polarized Bloch modes appear in pairs with the same $k_y$ but opposite $k_x$, as is required by the mirror symmetry of the slab that is not broken. The Hopf indices of the C-points would result in topological charges of Bloch modes being $+1/2$ and $-1/2$, at points of $\textbf{D}_{1,2}$ and $\textbf{D}_{3,4}$, respectively. Two-dimensional (2D) polarization distributions at indicated points around the pole are shown in Figs.~\ref{fig3}(b) and (d), where  polarization distributions show mirror symmetry at singularity pairs except for the opposite sign of $S_3$, due to the pseudo-scalar nature of chirality~\cite{jackson1962classical,BIRSS_1964}. All the obtained results shown here are consistent with Ref.~\cite{LIU_2019_ArXiv190401733Phys._Circularly}, where however such a connection with multipolar singularities had not been revealed, thus lacking a clear microscopic picture that we have rendered in this work.

Same principles are also applicable to  hexagonal crystal slabs [see inset of Fig.~\ref{fig4}(a)]. The dispersion curves [along $\phi_1=\arctan(k_y/k_x)=0.45^{\circ}$ and $\phi_2=180^{\circ}-\phi_1$] are shown in Fig.~\ref{fig4}(a). Point $\textbf{B}_{3}$ represents a $\Gamma$-point BIC with topological charge of $-2$, corresponding to a multipolar \textbf{V}-point of index $-2$~\cite{CHEN_2019__Singularities}. When the asymmetry is introduced ($\vartheta=0.05$), it would break into four \textbf{C}-points of the same index $-1/2$. The multipolar radiation along the open channel and thus $\Gamma$-point Bloch mode are both linearly-polarized ($S_3=0$). Similarly, those \textbf{C}-points can be engineered to coincide with the open channel at $\textbf{D}_{5, 6}$ ($k_xp/2\pi=\pm 0.0102$; $\phi=\phi_{1,2}$), generating a pair of Bloch modes of topological charge $-1/2$ [see Fig.~\ref{fig4}(b) for the corresponding 2D polarization distributions].  The other pair of Bloch modes not shown in Fig.~\ref{fig4} are presented in Ref.~\cite{Supplemental_Material}. In Figs.~\ref{fig3} and \ref{fig4}, we discuss only symmetry-protected $\Gamma$-point BICs, and similar effects can be observed with non-$\Gamma$-point accidental BICs (see Ref.~\cite{Supplemental_Material} for the demonstration).

In summary, we reexamine the general complex electromagnetic multipoles and show that, regardless of how different multipoles  are combined, the Poincar\'{e}-Hopf theorem  guarantees the existence of isolated singularities and the index sum to be $\mathbf{2}$. It is further revealed that, the formation of radiative circularly-polarized Bloch modes actually originate from  overlapping of multipolar \textbf{C}-points with open radiation channels, the Hopf indices of which decide the topological charges of those Bloch modes. Our work has explored both the local and global topological landscapes of electromagnetic multipoles, revealing their subtle connections with other core physical entities of line field singularities and Bloch modes, which can potentially incubate new concepts and applications for not only photonics, but also many other multipole-topology-related branches of physics and interdisciplinary studies.

\begin{acknowledgments}
We acknowledge the financial support from National Natural Science Foundation of China (Grants No. 11874026, No. 11404403 and No. 11874426), and the Outstanding Young Researcher Scheme of National University of Defense Technology. W. L. is grateful to  Y. S. Kivshar and M. V. Berry for invaluable correspondences.
\end{acknowledgments}


\vspace{30mm}

\setcounter{figure}{0}
\setcounter{table}{0}
\setcounter{equation}{0}

\makeatletter
\renewcommand{\thefigure}{S\@arabic\c@figure}
\makeatother

\makeatletter
\renewcommand{\theequation}{S\@arabic\c@equation}
\makeatother

\makeatletter
\renewcommand{\thetable}{S\@arabic\c@table}
\makeatother

\textbf{\textsc{Supplemental Material for``Line Singularities and Hopf Indices of Electromagnetic Multipoles"}}\\

\textit{This Supplemental Material includes the following eight sections: (\textbf{\uppercase\expandafter{\romannumeral1}}). Calculation of the normalized Stokes parameters; (\textbf{\uppercase\expandafter{\romannumeral2}}). Polarization distributions for multipoles of order $(l,m)$, with  $l=1\rightarrow3$ and $m=-l\rightarrow l$; (\textbf{\uppercase\expandafter{\romannumeral3}}). Stereographic projection from $1$-spinor $w$ to the associated Stokes vector $\mathbf{S}$; (\textbf{\uppercase\expandafter{\romannumeral4}}). More details of breaking \textbf{V}-points to \textbf{C}-points shown in Figs.~2(b)-(d); (\textbf{\uppercase\expandafter{\romannumeral5}}). Multipolar expansions and the radiation patterns for unit-cells of infinite periodic structures; (\textbf{\uppercase\expandafter{\romannumeral6}}). Multipolar compositions at the points indicated in Figs.~\ref{fig3} and \ref{fig4}; (\textbf{\uppercase\expandafter{\romannumeral7}}). Breaking accidental non-$\Gamma$-point BICs into circularly-polarized Bloch modes; (\textbf{\uppercase\expandafter{\romannumeral8}}). The other pair of circularly-polarized Bloch modes of the asymmetric hexagonal slab in Fig.~\ref{fig4}.}

\titleformat{\section}[hang]{\bfseries}{\thesection.\ }{0pt}{}
\renewcommand{\thesection}{\Roman{section}}
\renewcommand{\thesubsection}{\thesection.\Roman{subsection}}

\section{Calculation of the normalized Stokes parameters.}

At the point ($\theta$, $\phi$) on the momentum sphere, we can employ the complex parameter $\chi$ to characterize the polarization~\cite{Yariv2006_book_photonics}:
\begin{equation}
\label{chi}
\chi  = \exp (i\delta )\tan (\psi ) = \Phi /\Theta,
\end{equation}
where ($\Theta$, $\Phi$) are complex field components along ${{\bf{\hat e}}}_\theta$ and ${{\bf{\hat e}}}_\phi$, respectively. Then the normalized Stokes parameters $S_1$, $S_2$, and $S_3$ are~\cite{Yariv2006_book_photonics}:
\begin{equation}
{S_{1}=\cos 2 \psi}; ~~
{S_{2}=\sin 2 \psi \cos \delta};~~
{S_{3}=\sin 2 \psi \sin \delta}.
\end{equation}
These three parameters decide the unit Stokes vector $\mathbf{S}$ ($S_{1},S_{2},S_{3}$)  on the polarization Poincar\'{e} sphere. For the special case of zero tangent vector ($\Theta=\Phi=0$), we assign the value $\mathbf{S}=S_{1,2,3}=0$. Special care should be taken for the two points at the poles where neither ${{\bf{\hat e}}}_\theta$ nor ${{\bf{\hat e}}}_\phi$ are defined. The field components ($\Theta$, $\Phi$) and thus Stokes parameters at both poles can only be deduced from those of the neighbouring regions based on the continuity of the complex vectorial fields (refer to the Supplemental Material of Ref.~\cite{CHEN_2019__Singularities} for more details).

\section{Polarization distributions for multipoles of order $(l,m)$, with  $l=1\rightarrow3$ and $m=-l\rightarrow l$.}

\begin{figure}[tp]
\centerline{\includegraphics[width=8.9cm]{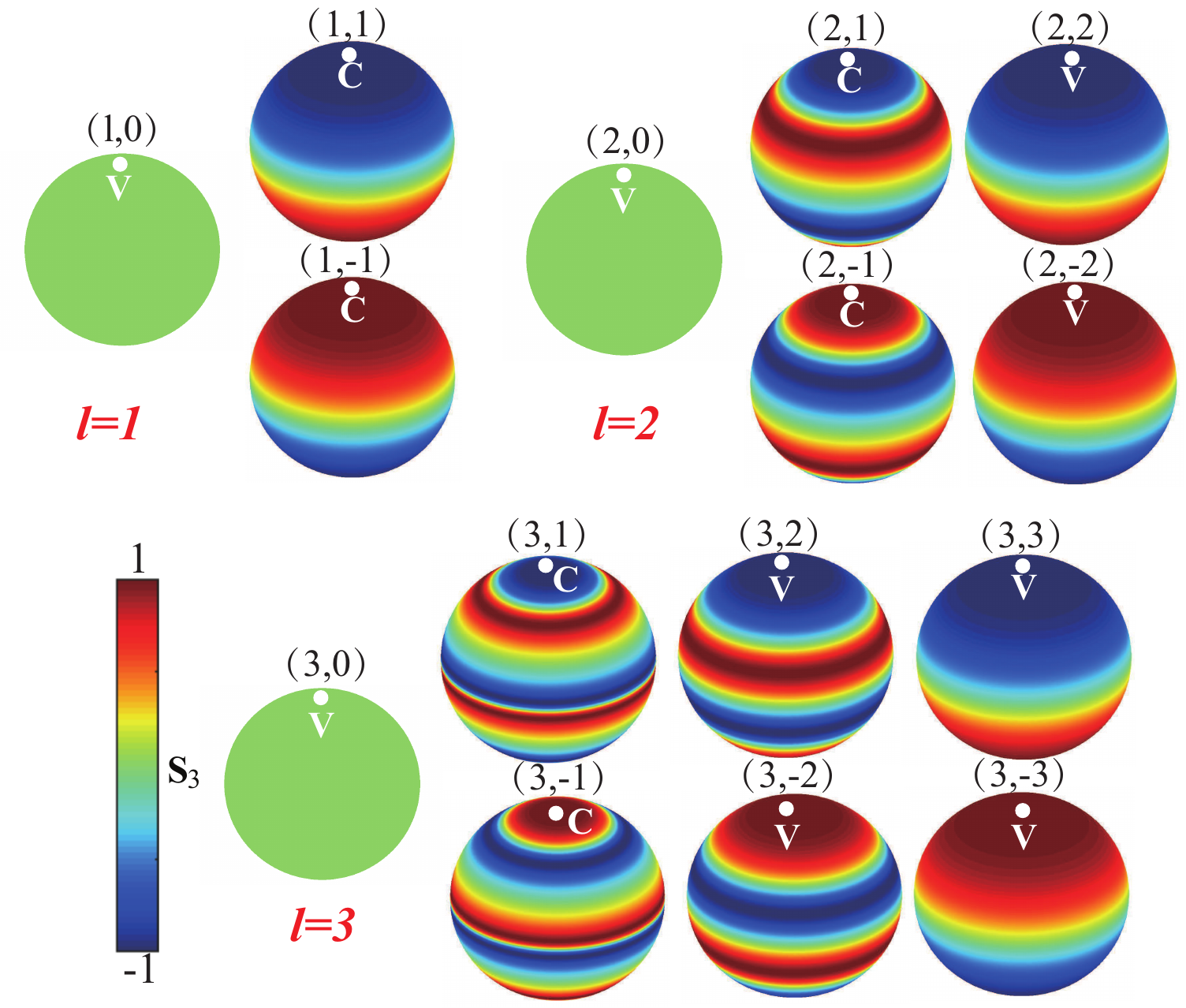}} \caption{\small Polarization distributions (across the momentum sphere in terms of $S_3$) for multipoles of order $(l,m)$,  with  $l=1\rightarrow3$ and $m=-l\rightarrow l$.}
\label{figs1}
\end{figure}

Figure~\ref{figs1} shows the polarization distributions (across the momentum sphere in terms of $S_3$) for multipoles of order $(l,m)$, with  $l=1\rightarrow3$ and $m=-l\rightarrow l$. Those distributions are consistent with two conclusions drawn in the main letter: (i) $S_3$ is an odd function of $m$: $S_3(\textbf{M}_{lm}; \textbf{N}_{lm})=-S_3(\textbf{M}_{l,-m}; \textbf{N}_{l,-m})$; (ii) the light field is linearly-polarized ($S_3=0$) all across the momentum sphere when $m=0$. Moreover, for all cases of $(l,m)$, there are only two isolated polarization singularities locating on the poles with Hopf index of $+1$. They are \textbf{C}-points for  $m=\pm 1$ and \textbf{V}-points for $m\neq\pm 1$.
\section{Stereographic projection from $1$-spinor $w$ to the associated Stokes vector $\mathbf{S}$.}
\begin{figure*}[tp]
\centerline{\includegraphics[width=13cm]{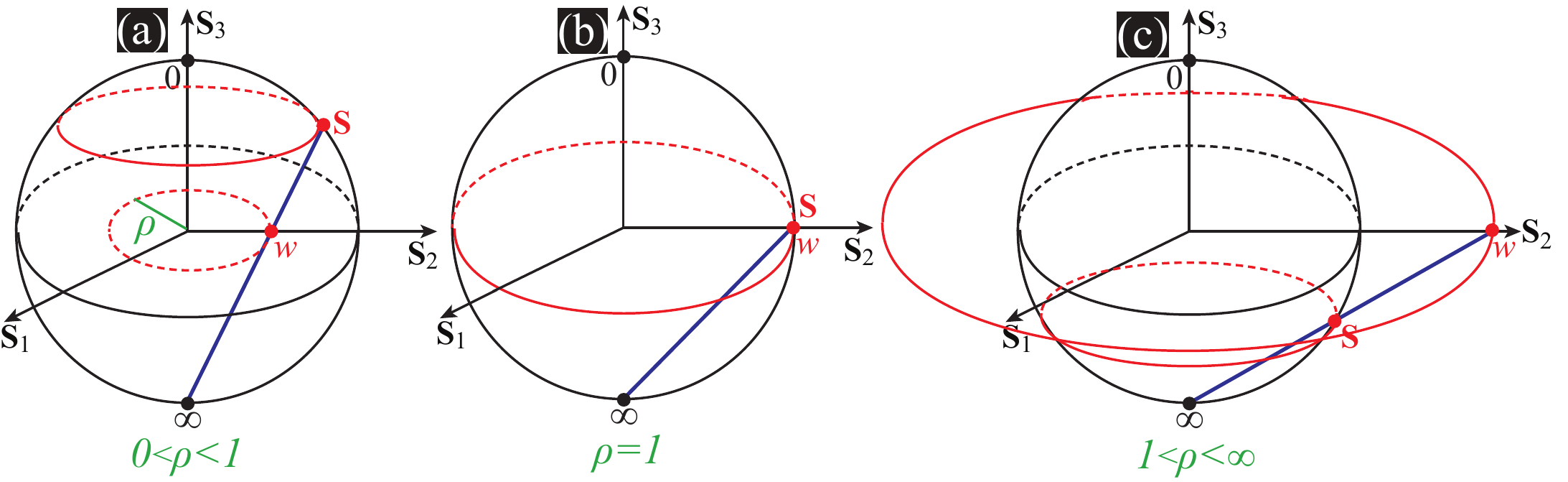}} \caption{\small Three cases of the stereographic projection ($0<\rho<1$; $\rho=1$; $1<\rho<\infty$)  from $w$-point on the $S_3=0$ plane to a Stokes vector on the Poincar\'{e} sphere.}
\label{figs2}
\end{figure*}
The complex parameter [$1$-spinor: $w=\rho(r_s)\exp(i\ell\phi_s)$] introduced with local circularly-polarized basis vectors in Eq. (4) of the main letter  can decide the polarization state and is directly related to the Stokes vector $\mathbf{S}$ through the stereographic projection~\cite{PENROSE_2004__Road,BERRYM.V._2003_Proceeding_optical,HLADIK_1999__Spinors}. Three cases of the stereographic projection ($0<\rho<1$; $\rho=1$; $1<\rho<\infty$) are shown in Fig.~\ref{figs2}. As is shown, the stereographic projection can be conducted through the following procedures: (i). Project the complex $w$ to the plane of $S_3=0$ (the corresponding $w$-plane of the Riemann sphere~\cite{PENROSE_2004__Road,BERRYM.V._2003_Proceeding_optical}) through $w=S_1+iS_2$; (ii). Connect the south pole and the $w$ point to form a line, and its crossing point with the Poincar\'{e} sphere decides the Stokes vector $\mathbf{S}$ (it is the south pole itself only for $\rho=\infty$); (iii). The stereographically projected vectors of two special cases of $\rho=0$ and $\rho=\infty$ are the north and south poles, which correspond to singularities (\textbf{C}-points) of left-handed and right-handed circularly-polarized light, respectively. For $0<\rho<\infty$, the associated vectors would locate on a latitude circle:  when $0<\rho<1$, the vectors are on the northern hemisphere and the corresponding states are left-handed elliptically-polarized [Fig.~\ref{figs2}(a)];  when $\rho=1$, the vectors are on the equator ($w$-point and $\mathbf{S}$-point coincide on the Poincar\'{e} sphere) and the corresponding states are linear-polarized [Fig.~\ref{figs2}(b)]; when $1<\rho<\infty$, the vectors are on the southern hemisphere and the corresponding states are right-handed elliptically-polarized [Fig.~\ref{figs2}(c)]. \\

As a circuit is traversed around the singularity ($r_s\rightarrow 0$; $\rho\rightarrow 0$ or $\infty$) with $\phi_s=0 \rightarrow 2\pi$, the corresponding Stokes vector \textbf{S} would trance out the corresponding complete latitude circle $|\ell|$ times, indicating $|\ell|/2$ rounds of $2\pi$ full rotations of the polarization lines along the circuit. The positive (negative) sign of $\ell$ means that rotation of the polarization line is in the same (opposite) sense as that the contour is traversed. As a result, the Hopf index of the singularity at the origin is  ${\mathbf{Ind}_h}(r_s=0)=\ell/2$.

\begin{figure}[tp]
\centerline{\includegraphics[width=7.5cm]{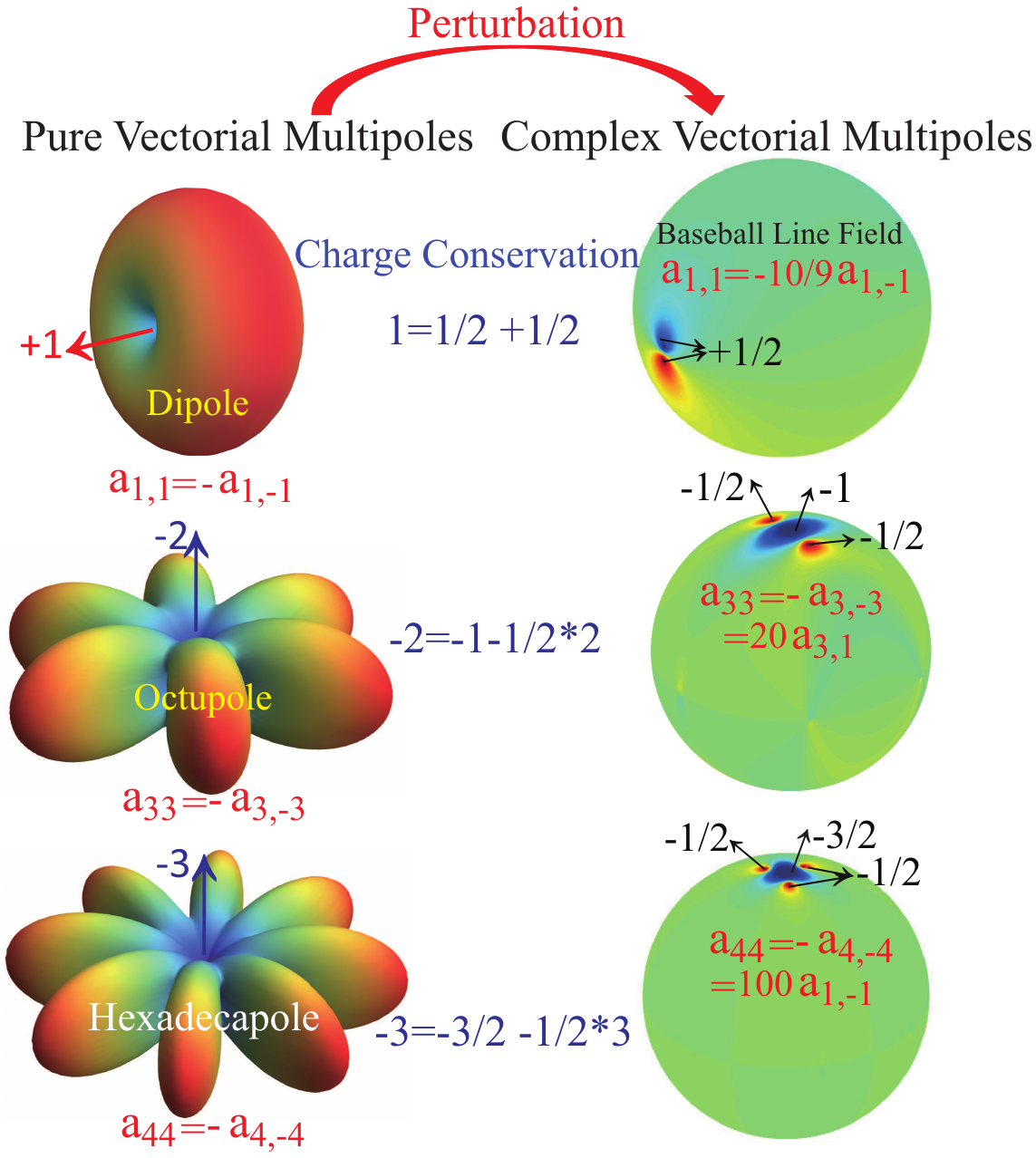}} \caption{\small Perturb the pure vectorial multipoles (left column; dipole, octupole and hexadecapole from top to bottom) to break the \textbf{V}-points (of Hopf indices $+1$, $-1$, and $-2$ respectively) into a series of \textbf{C}-points (right column) with the total topological charge (total index in the surrounding area) conserved. The left column shows the radiation patterns of pure vectorial multipoles and the right column shows the polarization distributions of the complex vectorial multipoles with perturbations.}
\label{figs3}
\end{figure}

\section{More details of breaking \textbf{V}-points to \textbf{C}-points shown in Figs.~2(b)-(d).}
In this section, we show more details of how to perturb the pure vectorial multipoles (those thoroughly  studied in Ref.~\cite{CHEN_2019__Singularities}) to break the \textbf{V}-points into a series of \textbf{C}-points [as shown in Figs.~2(b)-(d) of the main letter] with the total topological charge (total index) conserved. The results are summarized in Fig.~\ref{figs3}, where the left column shows the radiation patterns of pure vectorial multipoles (dipole, octupole and hexadecapole) and the right column shows the polarization distributions of the complex vectorial multipoles after perturbations. In the left column we do not show the polarization distributions, as for pure vectorial dipoles, it it is linearly-polarized everywhere across the momentum sphere with $S_3=0$. The singularities indicated are \textbf{V}-points (for pure vectors, all singularities are \textbf{V}-points) and \textbf{C}-points in the left and right column, respectively. The introduction of the $m=1$ perturbation term guarantees that the singularities at the poles are \textbf{C}-points. In Fig.~\ref{figs3} not all singularities are indicated, but it is easy to confirm that the index sum of all isolated singularities across the momentum sphere is always $\mathbf{2}$, as is required by the Poincar\'{e}-Hopf theorem.

\begin{figure*}[tp]
\centerline{\includegraphics[width=15cm]{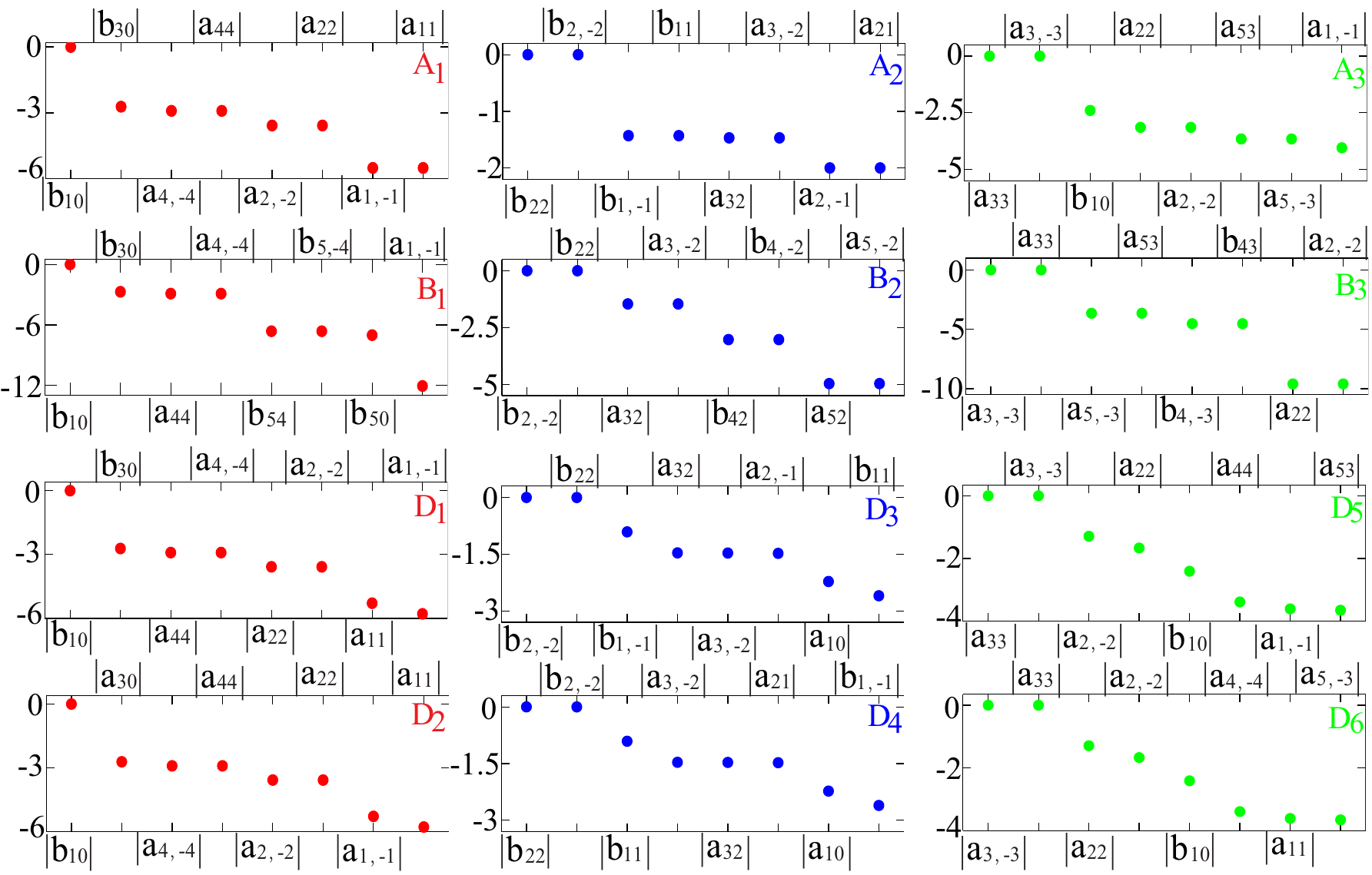}} \caption{\small Multipolar compositions (normalized magnitudes of expansion coefficients in logarithmic scale; only the dominantly contributing eight terms are shown) for the the points indicated in Figs.~\ref{fig3}(a) and (c),  and Fig.~\ref{fig4}(a), including $\textbf{A}_{1\rightarrow3}$ (linearly-polarized), $\textbf{B}_{1\rightarrow3}$ (BICs) and $\textbf{D}_{1\rightarrow6}$ (circularly polarized).}
\label{figs4}
\end{figure*}

\section{Multipolar expansions and the radiation patterns for unit-cells of infinite periodic structures.}
The approach to conduct multipolar expansions for unit-cells of periodic structures and to obtain the corresponding radiation patterns in the far field has already been discussed with detailed formulas in the Supplemental Material of Ref.~\cite{CHEN_2019__Singularities}. In this section, we reproduce the related information for the convenience of the readers of this work:\\

For finite particles or particles clusters, the radiated  fields along all directions can be (numerically) calculated, which can then be directly expanded to spherical harmonics of different orders with coefficients of $a_{lm}$ and $b_{lm}$. Consequently, the magnitudes and phases of all electric and magnetic multipoles would be obtained, as has been widely implemented in most previous studies~\cite{jahani_alldielectric_2016,KUZNETSOV_Science_optically_2016,LIU_2018_Opt.Express_Generalized}. This approach of radiation expansion (expansion of the radiated fields) can not be directly implemented for unit-cells of infinite periodic structures, as the radiation of infinite periodic structures is only known and allowed along certain directions that correspond to open diffraction channels.\\

For infinite periodic structures, we can carry out the multipolar expansion based on the near-field polarization currents $\bf{J(r)}$ of the unit-cell. The multipolar coefficients of $a_{lm}$ and $b_{lm}$ can be obtained through the following relations (when there are no effective magnetization currents)~\cite{jackson1962classical,GRAHN_NewJ.Phys._electromagnetic_2012}:
\begin{eqnarray}
&{a_{lm}} = {{{{\left( { - i} \right)}^{l - 1}}{k^2}\eta {Q_{lm}}} \over {{{\left[ {\pi \left( {2l + 1} \right)} \right]}^{{1 \over 2}}}}}\mathop{\int\!\!\!\int\!\!\!\int} \limits_\mathbf{cell} {{\bf{J}}\left( {\bf{r}} \right) \cdot {\mathbf{S}_{lm}}{d^3}r}; \label{current_expansion_1}\\
&{b_{lm}} = {{{{\left( { - i} \right)}^{l + 1}}{k^2}\eta {Q_{lm}}} \over {{{\left[ {\pi \left( {2l + 1} \right)} \right]}^{{1 \over 2}}}}}\mathop{\int\!\!\!\int\!\!\!\int}\limits_\mathbf{cell} {{\bf{J}}\left( {\bf{r}} \right) \cdot {\mathbf{T}_{lm}}{d^3}r},\label{current_expansion_2}
\end{eqnarray}  
where $\eta$ is the vacuum impendence; ${Q_{lm}} = {1 \over {{{\left[ {l\left( {l + 1} \right)} \right]}^{1/2}}}}{\left[ {{{2l + 1} \over {4\pi }}{{\left( {l - m} \right)!} \over {\left( {l + m} \right)!}}} \right]^{1/2}}$; the integration is conducted within the unit-cell region. As long as the electric field distributions ${\bf{E}}\left( {\bf{r}} \right)$ within the unit-cell is obtained (calculated through COMSOL Multiphysics in our work), the corresponding electric polarzation currents can be obtained as:
\begin{equation}
\label{electric current}
{\bf{J}}\left( {\bf{r}} \right) =  - i\omega {\varepsilon _0}\left[ {{\varepsilon _r}\left( {\bf{r}} \right) - {\varepsilon _b}} \right]{\bf{E}}\left( {\bf{r}} \right),
\end{equation}
where ${\varepsilon _r}\left( {\bf{r}} \right)$ denotes the permittivity distribution and ${\varepsilon _b}$ is the permittivity of the background medium. Vector functions $\mathbf{S}_{lm}$ and $\mathbf{T}_{lm}$ are:
\begin{eqnarray*}
\label{vector_function}
\begin{split}
{\mathbf{S}_{lm}} = \exp \left( { - im\phi } \right)\left[ {{\Pi _l}\left( {kr} \right) + {\Pi _l}^{\prime \prime }\left( {kr} \right)} \right]P_l^m\left( {\cos \theta } \right){\hat{\mathbf{e}}_r} \\
+ \exp \left( { - im\phi } \right){{{\Pi _l}^\prime \left( {kr} \right)} \over {kr}}\left[ {{\tau _{lm}}\left( \theta  \right){\hat{\mathbf{e}}_\theta } - i{\pi _{lm}}\left( \theta  \right){\hat{\mathbf{e}}_\phi }} \right]
\end{split};\\
{\mathbf{T}_{lm}} = \exp \left( { - im\phi } \right){j_l}\left( {kr} \right)\left[ {i{\pi _{lm}}\left( \theta  \right)\hat{\mathbf{e}}_ \theta  + {\tau _{lm}}\left( \theta  \right)\hat{\mathbf{e}}_ \phi } \right],
\end{eqnarray*}
for which (also defined in the main letter) ${\tau _{lm}}\left( \theta  \right) = {d \over {d\theta }}P_l^m\left( {\cos \theta } \right)$, ${\pi _{lm}}\left( \theta  \right) = {m \over {\sin \theta }}P_l^m\left( {\cos \theta } \right)$, and ${\Pi _l}\left( {kr} \right) = kr{j_l}\left( {kr} \right)$ is Riccati-Bessel function [$j_l(kr)$ is the spherical Bessel function of the first kind]~\cite{BRONSHTEIN_2007__Handbook}.\\

With the expansion coefficients $a_{lm}$ and $b_{lm}$  obtained through Eqs.~(\ref{current_expansion_1}) and (\ref{current_expansion_2}), the radiated fields from the unit-cell can be reconstructed as follows:
\begin{equation}
\label{radiated_fields}
{{\bf{E}}_{\rm{rad}}}\left( {r,\theta ,\phi } \right) = \sum\limits_{l = 1}^\infty  {\sum\limits_{m =  - l}^l {{E_{lm}}\left[ {{a_{lm}}{{\bf{{N}}}_{lm}} + {b_{lm}}{{\bf{{M}}}_{lm}}} \right]} },
\end{equation}
where ${E_{lm}} = {{{i^{l + 1}}\left( {2l + 1} \right)} \over 2}\sqrt {{{\left( {l - m} \right)!} \over {l\left( {l + 1} \right)\left( {l + m} \right)!}}}$; and vector functions $\bf{{N}}$$_{lm}$ and $\bf{{M}}$$_{lm}$ are:
\begin{eqnarray}
\label{vector_function_radiation}
\begin{split}
{{{\bf{{N}}}}_{lm}}{\rm{ = }}\left[ {{\tau _{lm}}\left( {\cos \theta } \right){{{\bf{\hat e}}}_\theta } + i{\pi _{lm}}\left( {\cos \theta } \right){{{\bf{\hat e}}}_\phi }} \right]{{{{\left[ {kr{h_l}^{\left( 1 \right)}\left( {kr} \right)} \right]}^\prime }} \over {kr}}\nonumber\\
+ \exp \left( {im\phi } \right){{{\bf{\hat e}}}_r}l\left( {l + 1} \right)P_l^m\left( {\cos \theta } \right){{{h_l}^{\left( 1 \right)}\left( {kr} \right)} \over {kr}}\exp \left( {im\phi } \right);
\end{split}\\
\begin{split}
{\mathbf{{M}}_{lm}} = \left[ {i{\pi _{lm}}\left( {\cos \theta } \right){{{\bf{\hat e}}}_\theta } - {\tau _{lm}}\left( {\cos \theta } \right)
{{{\bf{\hat e}}}_\phi }} \right]\nonumber\\
{h_l}^{\left( 1 \right)}\left( {kr} \right)\exp \left( {im\phi } \right),
\end{split}
\end{eqnarray}
where ${h_l}^{\left( 1 \right)}\left( {kr} \right)$ is the spherical Hankel function of the first kind~\cite{BRONSHTEIN_2007__Handbook}, and here we have provided the full expression of ${\mathbf{{N}}_{lm}}$ that includes also the radial component.\\

\section{Multipolar compositions at the points indicated in Figs.~\ref{fig3} and \ref{fig4}.}

Specific multipolar compositions (normalized magnitudes of expansion coefficients in logarithmic scale; only the dominantly contributing eight terms are shown) for the points indicated in Figs.~\ref{fig3} and \ref{fig4}, including
$\textbf{A}_{1\rightarrow3}$, $\textbf{B}_{1\rightarrow3}$ and $\textbf{D}_{1\rightarrow6}$.

\section{Breaking accidental non-$\Gamma$-point BICs into circularly-polarized Bloch modes.}

\begin{figure}[tp]
\centerline{\includegraphics[width=8.8cm]{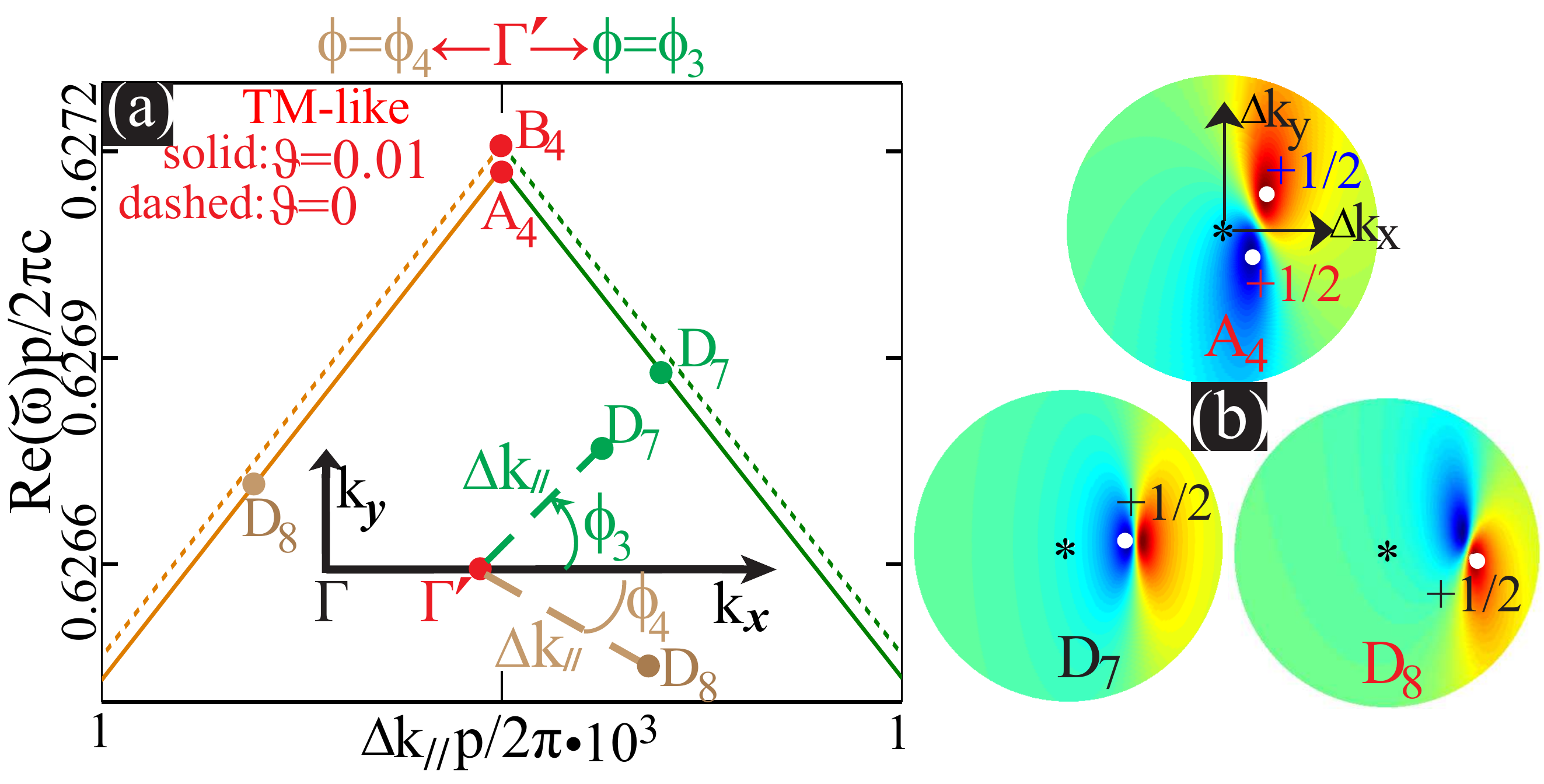}} \caption{\small (a): Dispersion curves (both symmetric dashed and assymetric solid) for square photonic crystal slabs and four Bloch modes are indicated: $\textbf{A}_{4}$ (elliptically-polarized); $\textbf{B}_{4}$ (BIC);  $\textbf{D}_{7,8}$ (circularly-polarized). More definitions of the parameters are shown in the inset, where for symmetric case ($\vartheta=0$), the BIC (denoted by point $\textbf{B}_4$) locates at $\Gamma'$ point ($k_xp/2\pi=0.299$; $k_y=0$). The corresponding 2D polarization distributions (around the BIC-$\Gamma'$ point denoted by asterisks) are shown in (b), with \textbf{C}-points (denoted by dots) and their indices specified.}
\label{figs5}
\end{figure}

In Figs.~\ref{fig3} and \ref{fig4} of the main letter, we have confined our discussions to symmetry-protected $\Gamma$-point BICs and how they can evolve into pairs of circularly-polarized Bloch modes by symmetry breaking. Here we show that the same principle can be applied to non-$\Gamma$-point BICs, from which in a similar way circularly-polarized Bloch modes can be also obtained. Here  we break the symmetry of a photonic crystal slab with square lattices of circular air holes by filling in part of the holes (parameters are the same as those in Fig.~\ref{fig3} of the main letter). The dispersion curves [both symmetric dashed curves  ($\vartheta=0$) and asymmetric solid curves ($\vartheta=0$)] of the TM-like bands (along $\phi_3=7.13^{\circ}$ and $\phi_4=-4.76^{\circ}$) are shown in Fig.~\ref{figs5}(a). For symmetric case, the BIC [denoted by point $\textbf{B}_4$ in Fig.~\ref{figs5}(a)] locates at $\Gamma'$ point ($k_xp/2\pi=0.299$; $k_y=0$).
As has been revealed in~\cite{CHEN_2019__Singularities}, the topological charge of this BIC is $+1$, which corresponds to the \textbf{V}-point of the multipolar radiation (along the corresponding open channel $k_x=k_y=0$) with Hopf index of $+1$.\\

When the symmetry is broken ($\vartheta=0.01$),  similar to the scenarios discussed in Fig.~2 in the main letter, at $\textbf{A}_{4}$ [indicated on the dispersion curve for the asymmetric case in Fig.~\ref{figs5}(a)] the \textbf{V}-point of the radiation will break into a pair of singularities of indices $+1/2$ with opposite handedness [$\textbf{A}_{4}$ in Fig.~\ref{figs5}(b)]. More definitions of the parameters related to such breaking is shown in the inset of Fig.~\ref{figs5}(a). In contrast to the $\Gamma$-point BIC where the $\Gamma$-point Bloch modes would be linearly-polarized after symmetry breaking, here the BIC would evolve to a Bloch mode of elliptic polarization of $S_3\neq 0$ [$\textbf{A}_{4}$ in Fig.~\ref{figs5}(b)], as there is no more mirror symmetry at non-$\Gamma$ point. When the \textbf{C}-points are made to overlap with the open diffraction channels, which is satisfied at the points $\textbf{D}_{7,8}$ ($\Delta k_{\|}p/2\pi=0.0004,~0.0006$ and $\phi=\phi_3,~\phi_4$, respectively; $\Delta k_{\|}=\sqrt{\Delta k_{x}^2+\Delta k_{y}^2}$) indicated in Fig.~\ref{figs5}(a). The Hopf indices of the C-points would result in topological charges of Bloch modes being $+1/2$ at points $\textbf{D}_{7,8}$ [refer to Figs.~\ref{figs5}(b) for the corresponding polarization distributions around the BIC position of $k_xp/2\pi=0.99$;~$k_y=0$ that are represented by asterisks].

\section{The other pair of circularly-polarized Bloch modes of the asymmetric hexagonal slab in Fig.~\ref{fig4}.}

\begin{figure}[htp]
\centerline{\includegraphics[width=8.5cm]{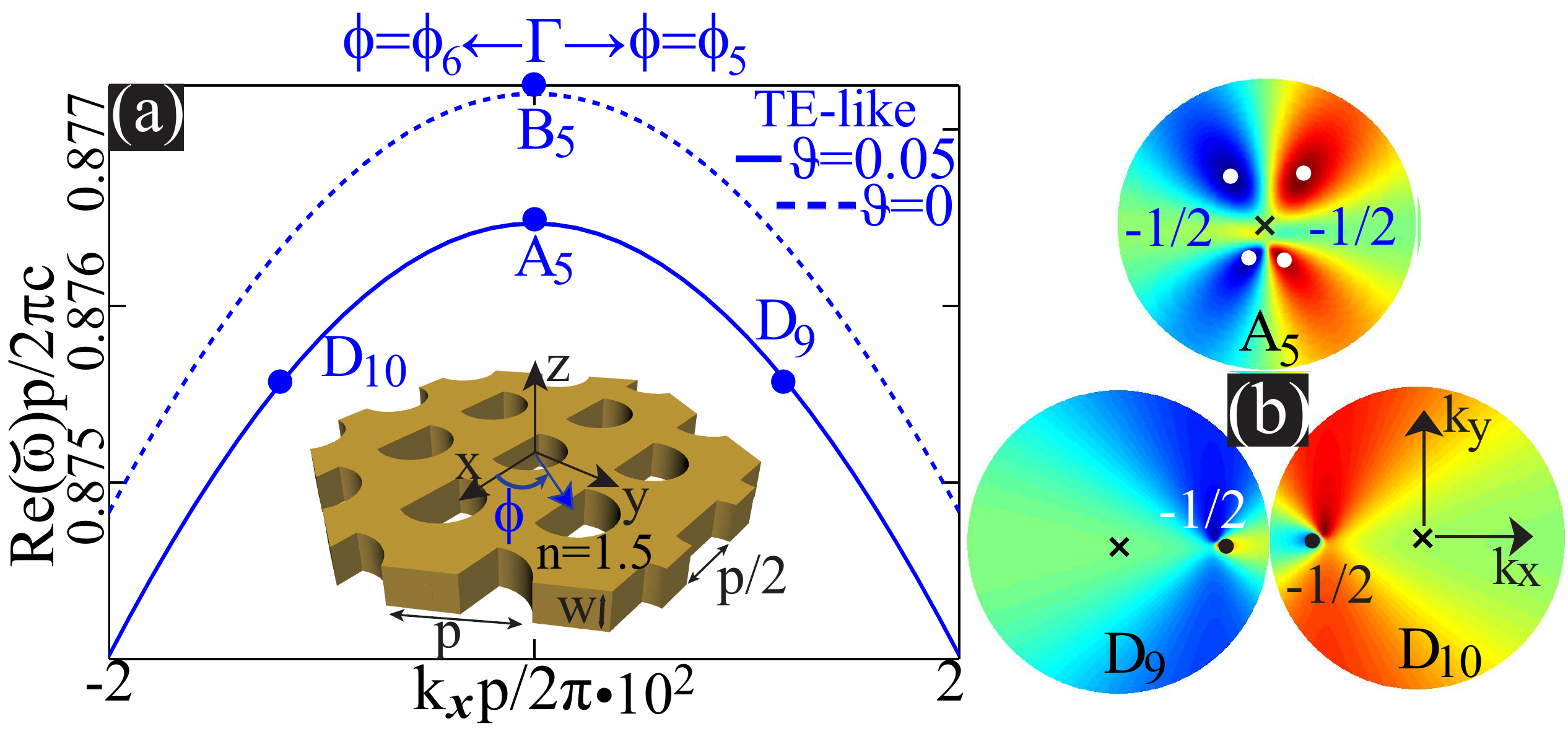}} \caption{\small (a): Dispersion curves (along $\phi_5=-1.28^{\circ}$ and $\phi_6=-180^{\circ}-\phi_5$) for the hexagonal photonic crystal slab studied in Fig.~\ref{fig4}. Four Bloch modes are indicated: $\textbf{A}_{5}$ (linearly-polarized); $\textbf{B}_{5}$ (BIC);  $\textbf{D}_{9,10}$ (circularly-polarized). The corresponding 2D polarization distributions (around the pole of $k_x=k_y=0$ denoted by crosses)  are shown in (b), with \textbf{C}-points (denoted by dots) and their indices specified.}
\label{figs6}
\end{figure}

As is shown in Fig.~\ref{fig4} of the main letter, when the asymmetry is introduced into the hexagonal photonic crystal slab ($\vartheta=0.05$), four \textbf{C}-points of the same index $-1/2$ [see Fig.~\ref{fig4}(b) of the main letter] are obtained by breaking the \textbf{V}-point of index $-2$ that is associated with the BIC state of an equivalent topological charge of $-2$ [indicated by $\textbf{B}_{3}$ in Fig.~\ref{fig4}(a)]. Those four singularities can be engineered to overlap with the open diffraction channels, producing the same number of four radiative circularly-polarized Bloch modes of topological charge of $-1/2$. One pair of those Bloch modes (located on $\phi_1=0.45^{\circ}$ and $\phi_2=180^{\circ}-\phi_1$ dispersion branches) are presented in Fig.~\ref{fig4} [indicated respectively by points $\textbf{D}_{5,6}$ in Fig.~\ref{fig4}(a)]. The other pair of circularly-polarized Bloch modes locates on the dispersion branches of $\phi_5=-1.28^{\circ}$ and $\phi_6=-180^{\circ}-\phi_5$, indicated by $\textbf{D}_{9,10}$ ($k_xp/2\pi=\pm 0.0103$; $\phi=\phi_{5,6}$) in Fig.~\ref{figs6}(a), respectively. The multipolar radiation along the open channel and thus $\Gamma$-point Bloch mode are both linearly-polarized ($S_3=0$). Both Bloch modes are of topological charge $-1/2$, which agrees with the index of the overlapping singularities. The corresponding 2D polarization distributions around the pole (denoted by crosses) are shown Fig.~\ref{figs6}(b).

\end{document}